\documentclass[letterpaper, 10 pt, conference]{ieeeconf}
\IEEEoverridecommandlockouts
\usepackage{cite}
\usepackage{amsmath,amssymb,amsfonts}
\usepackage{algorithmic}
\usepackage{graphicx}
\usepackage{textcomp}
\usepackage{xcolor}
\usepackage{mathtools}
\usepackage{ntheorem}
\usepackage{epstopdf}
\usepackage{caption}
\usepackage{subcaption}
\usepackage{setspace}
\usepackage{subcaption}
\usepackage{dsfont}
\usepackage{cleveref}

\makeatletter
\renewcommand*\env@matrix[1][*\c@MaxMatrixCols c]{%
  \hskip -\arraycolsep
  \let\@ifnextchar\new@ifnextchar
  \array{#1}}
\makeatother

\newcommand{\Real}{{\mathds R}} 
\newcommand{\Nat}{{\mathds N}} 

\newtheorem{definition}{Definition}{}
\newtheorem{corollary}{Corollary}{}
{}
{}
\newtheorem{theorem}{Theorem}{}
\newtheorem{remark}{Remark}{}
\newtheorem{lemma}{Lemma}{}
{}

\def\BibTeX{{\rm B\kern-.05em{\sc i\kern-.025em b}\kern-.08em
    T\kern-.1667em\lower.7ex\hbox{E}\kern-.125emX}}
\begin{document}

\title{\LARGE \bf
Finite Horizon Privacy of Stochastic Dynamical Systems:\\ A Synthesis Framework for Dependent Gaussian Mechanisms
}

\author{Haleh Hayati, Carlos Murguia, Nathan van de Wouw
\thanks{Haleh Hayati, Carlos Murguia, and Nathan van de Wouw are with the Department of Mechanical Engineering, Dynamics and Control Group, Eindhoven University of Technology, The Netherlands. Emails: \& h.hayati@tue.nl, \& c.g.murguia@tue.nl, \& n.v.d.wouw@tue.nl.}
}
\maketitle
\begin{abstract}
We address the problem of synthesizing distorting mechanisms that maximize privacy of stochastic dynamical systems. Information about the system state is obtained through sensor measurements. This data is transmitted to a remote station through an unsecured/public communication network. We aim to keep part of the system state private (a private output); however, because the network is unsecured, adversaries might access sensor data and input signals, which can be used to estimate private outputs. To prevent an accurate estimation, we pass sensor data and input signals through a distorting  (privacy-preserving) mechanism  before transmission, and send the distorted data to the trusted user. These mechanisms consist of a coordinate transformation and additive dependent Gaussian vectors. We formulate the synthesis of the distorting mechanisms as a convex program, where we minimize the mutual information (our privacy metric) between an arbitrarily large sequence of private outputs and the disclosed distorted data for desired distortion levels -- how different actual and distorted data are allowed to be.
\end{abstract}

\section{Introduction}
In a hyperconnected world, scientific and technological advances have led to an overwhelming amount of user data collected and processed by hundreds of companies over public networks. Companies mine this data to provide personalized services. These technologies have, however, come with the price of an extensive loss of privacy in society. Depending on their resources, adversaries can infer critical (private) information about system operations from public data available on the internet and/or unsecured servers and networks. This is why researchers from different fields (e.g., computer science, information theory, and control theory) have been attracted to the broad research area of privacy and security of Cyber-Physical Systems (CPSs), \cite{Farokhi1}\nocite{Farokhi2}\nocite{Pappas}\nocite{Jerome1}\nocite{Takashi_1}\nocite{Takashi_3}-\nocite{chaper_privacy_chaos}\hspace{-.1mm}\cite{Carlos_Iman1}.\\
In most engineering applications, information about the state of systems, say $X$, is obtained through sensor measurements and then sent to a remote station through communication networks for signal processing and decision-making purposes. If the communication network is public/unsecured and/or the remote station is untrustworthy, adversaries might access and estimate the system state. \\
A common technique to avoid an accurate estimation is the use of additive random vectors to distort disclosed data. In the context of privacy of databases, a popular approach is differential privacy \cite{Jerome1,Dwork}, where random noise is added to the response of queries so that private information stored in the database cannot be inferred. In general, if the data to be kept private follows continuous probability distributions, the problem of finding the optimal additive noise to maximize privacy is hard to solve. This problem has been addressed by assuming the data to be kept private is deterministic \cite{Farokhi1,SORIA,Geng}. However, in a Cyber-Physical-Systems context, the inherent system dynamics and unavoidable system and sensor noise lead to stochastic non-stationary data and thus, existing tools do not fit this setting. More recently, the authors in \cite{Carlos_Iman1,murguia2021privacy} have proposed a framework for synthesizing optimal transition probabilities to maximize privacy for a class of quantized CPSs with discrete multivariate probability distributions. Even though this framework is general and leads to distorting mechanisms with arbitrary distributions, the computational complexity induced by exploring all the possible transition probabilities from private to the disclosed distorted data is high and increases exponentially with the alphabet of the private data. In \cite{Farokhi1}, the authors neglect quantization and work directly with dynamical systems driven by continuous (Gaussian) disturbances. They prove that, in the unconstrained additive noise case (so distortion is not considered), the optimal noise distribution minimizing the Fisher information (their privacy metric) is Gaussian. This observation has also been made in \cite{Cedric} where mutual information is used as privacy metric.\\
Motivated by these results, in this manuscript, we present an optimization-based framework for synthesising privacy-preserving Gaussian mechanisms that maximize privacy but keep distortion bounded. We use additive Gaussian dependent vectors as distorting mechanisms to maximize privacy. We pass sensor data and input signals through these distorting mechanisms before transmission and send the distorted data to the remote station instead. These mechanisms consist of a coordinate transformation and additive dependent Gaussian vectors that are designed to hide (as much as possible) the private parts of the state $S$ -- a desired private output modeled as some linear function of the system state, $S=DX$, for some deterministic matrix $D$.\\
 Note, however, that it is not desired to overly distort the original data. When designing the additive Gaussian vectors, we need to take into account the trade-off between \emph{privacy} and \emph{distortion}. As \emph{distortion metric}, we use a general \emph{weighted mean squared error} between the original and distorted data. Weighting matrices are used to fine tune the desired distortion at different channels and/or to model different applications of the distorted data at the remote station. In this manuscript, we follow an information-theoretic approach to privacy. As \emph{privacy metric}, we propose a combination of \emph{mutual information} and \emph{entropy} \cite{Cover} between disclosed and private data. In particular, we aim at minimizing the mutual information $I[S;Z]$, between the private output $S$ and the disclosed randomized sensor data $Z$, while maximizing the entropy $h[H]$ of an additive Gaussian vector $H$ we use to distort input signals, over a finite time window, for \emph{desired levels of distortion} -- how different actual and distorted data are allowed to be. As we prove in this manuscript, we can cast the problem of finding the optimal Gaussian distributions and change of coordinates as a constrained convex optimization problem.\\
\textbf{Notation:} The symbol $\Real$ stands for the real numbers, $\Real_{>0}$($\Real_{\geq 0}$) denotes the set of positive (non-negative) real numbers. The symbol $\Nat$ stands for the set of natural numbers. The Euclidian norm in $\Real^n$ is denoted by $||X||$, $||X||^2=X^{\top}X$, where $^{\top}$ denotes transposition. The $n \times n$ identity matrix is denoted by $I_n$ or simply $I$ if $n$ is clear from the context. Similarly, $n \times m$ matrices composed of only ones and only zeros are denoted by $\mathbf{1}_{n \times m}$ and $\mathbf{0}_{n \times m}$, respectively, or simply $\mathbf{1}$ and $\mathbf{0}$ when their dimensions are clear. For positive definite (semidefinite) matrices, we use the notation $P>0$ ($P \geq 0$). For any two matrices $A$ and $B$, the notation $A \otimes B$ (the Kronecker product) stands for the matrix composed of submatrices $A_{ij}B$ , where $A_{ij}$, $i,j=1,...,n$, stands for the $ij-$th entry of the $n \times n$ matrix $A$. The notation $X \sim \mathcal{N}[\mu,\Sigma^X]$ means that $X \in \Real^{n}$ is a normally distributed random vector with mean $E[X] = \mu \in \Real^{n}$ and covariance matrix $E[(X-\mu)(X-\mu)^T] = \Sigma^X \in \Real^{n \times n}$, where $E[a]$ denotes the expected value of the random vector $a$. Finite sequences of vectors are written as $X^K := (X(1)^{\top},\ldots,X(K)^{\top})^{\top} \in \Real^{Kn}$, $X(i) \in \Real^{n}$, $i=1,...,K$, and $n,K \in \Nat$. To avoid confusion, we denote powers of matrices as $(A)^{K} = A \cdots A$ ($K$ times) for $K > 0$, $(A)^{0} = I$, and $(A)^{K} = \mathbf{0}$ for $K < 0$. The operators $\log[\cdot]$, $\det[\cdot]$, and $\text{tr}[\cdot]$ stand for logarithm base two, determinant, and trace, respectively.

\section{Preliminaries}
In this section, we present some definitions and preliminary results needed for the subsequent sections.

\begin{definition} [MMSE Estimator \cite{huemer2017component}] \label{linearMMSE}
Let $X$ and $Y$ be two jointly Gaussian random vectors. The Minimum Mean Square Error (MMSE) estimate of $X$ given $Y$ is given by\emph{:}
\begin{eqnarray}
\hat X = \Sigma ^{XY} {\Sigma^Y}^{-1} \left(Y - E(Y) \right) + E(X),
\end{eqnarray}
where $\Sigma ^{XY}$ denotes the cross-covariance matrix between $X$ and $Y$ and $\Sigma^Y$ is the covariance matrix of $Y$.
\end{definition}

\begin{definition}[Differential Entropy \cite{Cover}]\label{entropy}
Let $S \in \Real^n$ with $S \sim \mathcal{N}[\mu^S,\Sigma^S]$. Its differential entropy, $h(S)$, can be written in terms of its covariance matrix as\emph{:}
\begin{eqnarray}
h[S] = \frac{1}{2}\log \det \left( \Sigma^S \right) + \frac{n}{2} + \frac{n}{2}\log(2\pi). \label{entropyCov}
\end{eqnarray}
\end{definition}
Entropy is a measure of the average uncertainty in a random vector. We use base two $\log(\cdot)$, so entropy is given in bits.

\begin{definition}[Mutual Information \cite{Cover}]\label{mutual_info}
Let $S$ and $Z$ be two jointly distributed continuous random vectors with joint entropy $h[S,Z]$ and marginal entropies $h[S]$ and $h[Z]$. Their mutual information, $I[S;Z]$, is given as\emph{:}
\begin{eqnarray}
I[S;Z] = h[S] + h[Z] - h[S,Z].\label{mutinf}
\end{eqnarray}
\end{definition}
Mutual information between two jointly distributed vectors is a measure of the statistical dependence between them.

\section{Problem Formulation}
\subsection{System Description}
We consider discrete-time stochastic systems of the form:
\begin{eqnarray}\label{eq1}
\left\{ \begin{split}
X(k+1) &= AX(k) + BU(k) +  T(k),\\
Y(k) &= CX(k) + W(k),\\
S(k) &= DX(k),
\end{split} \right.
\end{eqnarray}
with time-index $k \in \Nat$, state $X \in {\mathbb{R}^{{n_x}}}$, measurable output $Y \in {\mathbb{R}^{{n_y}}}$, \emph{known} input $U \in {\mathbb{R}^{{n_u}}}$, private performance output $S \in {\mathbb{R}^{{n_s}}}$, and matrices $(A,B,C,D)$ of appropriate dimensions, ${n_x}, {n_y}, {n_u}, {n_s} \in \mathbb{N}$. Matrix $D$ is full row rank. The state perturbation $T$ and the output perturbation $W$ are multivariate i.i.d. Gaussian processes with zero mean and covariance matrices ${\Sigma ^T}>0$ and ${\Sigma ^W}>0$, respectively. The initial state $X(1)$ is assumed to be a Gaussian random vector with $E[X(1)]=\mu^X_1 \in \mathbb{R}^{n_x}$ and covariance matrix $\Sigma^X_1 = E[(X(1)-\mu^X_1)(X(1)-\mu^X_1)^{\top}] \in \mathbb{R}^{n_x \times n_x}$, $\Sigma^X_1 > 0$. Processes $T$ and $W$ and the initial condition $X(1)$ are mutually independent. We assume that matrices (vectors) $(A,B,C,D,\Sigma^X_1,\mu^X_1,\Sigma^T,\Sigma^W)$ and the input signal $U\left( k \right)$ are known for all $k$.\\
We aim to prevent adversaries from estimating the private output $S(k)$, accurately. To this end, we randomize measurements $Y(k)$ and input signals $U(k)$ before transmission and send the corrupted data to the remote station instead. The idea is to randomize $Y(k)$ and $U(k)$ as
\begin{equation}\label{eq2}
\left\{
\begin{array}{ll}
  Z(k) = G(k)Y(k) + V(k),\\[2mm]
  R(k) = U(k) + H(k),
\end{array}
\right.
\end{equation}
for some time-varying  transformation $G(k) \in {\mathbb{R}^{{n_y} \times {n_y}}}$ and dependent Gaussian processes, $V(k) \sim \mathcal{N}[\mathbf{0},\Sigma^V(k)]$ and $H(k) \sim \mathcal{N}[\mathbf{0},\Sigma^H(k)]$.
\begin{figure*}
  \centering
  \includegraphics[width=0.7\textwidth]{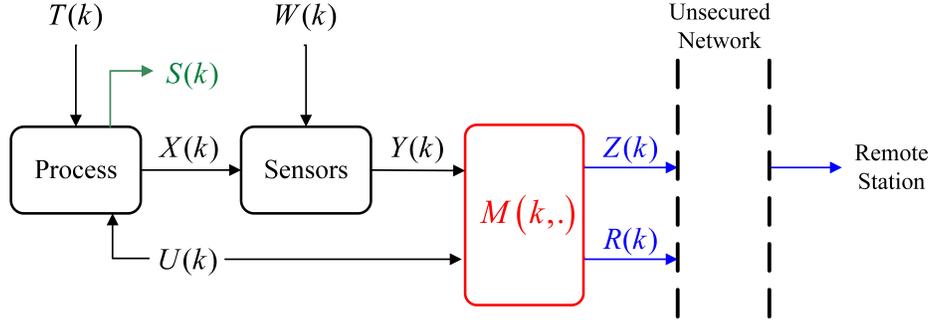}
    \caption{System configuration.}
    \label{fig2}
\end{figure*}
The randomized vectors $Z(k)$ and $R(k)$ are transmitted over an unsecured communication network to a remote station, see Figure \ref{fig2}. We seek to synthesize the sequences $G(k)$, $\Sigma^V(k)$, and $\Sigma^H(k)$, $k \in \mathcal{K}:= \{1,\ldots,K\}$, to make inference of the sequence of private outputs, $S(k)$, as `hard' as possible from the disclosed data, $(Z(k),R(k))$. In what follows, we introduce the adversarial model we seek to defend against.

\subsection{Adversarial Capabilities}
We consider worst-case adversaries that eavesdrop data at the communication network and/or the remote station. They do not only have access to all distorted sensor measurements $Z(k)$ and distorted input signal $R(k)$, but also have prior knowledge of the dynamics and the stochastic properties of the system, i.e., matrices  $(A,B,C,D,\Sigma^X_1,\mu^X_1,\Sigma^T,\Sigma^W)$ are known by the adversary. Moreover, the adversary also knows the means and covariance matrices $(\mu^Z(k),\mu^R(k),\Sigma^Z(k),\Sigma^R(k))$ as these can be estimated from the disclosed data $(Z(k),R(k))$. We assume that the adversary uses a linear MMSE estimator (see Definition 1) to reconstruct $S(k)$, which, for jointly Gaussian vectors, produces the best estimation performance among all unbiased estimators \cite{huemer2017component}. In practice, actual adversaries would typically not have all the capabilities that we assume here. However, if we maximize privacy under such worst-case adversaries, we ensure that adversaries with less capabilities perform even worse (or equal at most).

\subsection{Metrics and Problem Formulation}
For a given time horizon $K \in \mathbb{N}$, the aim of our privacy scheme is to make inference of the sequence of private vectors, ${S^K} = (S(1)^\top,...,S(K)^\top)^\top$, from the distorted disclosed sequences, ${Z^K} = (Z(1)^\top,...,Z(K)^\top)^\top$ and \linebreak ${R^{K}} = (R(1)^\top,...,R(K)^\top)^\top$, as hard as possible without distorting $(Y^K,U^{K})$, excessively. That is, we do not want to make $(Y^K,U^{K})$ and $(Z^K,R^{K})$ overly different. Hence, when designing the distorting variables $(G(k),\Sigma^V(k),\Sigma^H(k))$, we need to consider the \emph{trade-off between privacy and distortion}.\\
As distortion metric, we use the weighted mean squared errors between the original and distorted data, i.e., $E[||{W_Y}(Z^K - Y^K)||^2]$ and $E[||{W_U}(R^K - U^K)||^2]$, for some given weighting matrices of appropiate dimensions. Matrices $W_Y,W_U$ are used to fine-tune the desired distortion at different channels and/or to model different applications of the distorted data at the remote station. Arguably, for the class of linear systems considered in this manuscript, and most applications at the remote station (for this class), performance degradation induced by the privacy mechanism can be written (or upper bounded) in terms of the proposed weighted mean squared errors.\\
An intuitive candidate to use as privacy metric is the mutual information between private and disclosed data, i.e., $I[S^K;Z^K,R^{K}]$. However, because the input sequence $U^{K}$ is deterministic and $V(k)$ and $H(k)$ are independent, it is easy to verify that $I[S^K;Z^K,R^{K}] = I[S^K;Z^K]$. That is, in the proposed setting, the randomized input data $R^{K}$ does not affect $I[S^K;Z^K,R^{K}]$ at all. To overcome this obstacle, we add to $I[S^K;Z^K]$ the negative differential entropy, $h[U^{K}-R^{K}] = h[H^{K}]$, to capture the uncertainty between original and disclosed input data. That is, we propose $I[S^K;Z^K] - h[H^{K}]$ as privacy metric.\\
Summarizing the above discussion, we aim at minimizing $I[S^K;Z^K] - h[H^{K}]$ subject to the weighted second moment constraints $E[||{W_Y}(Z^K - Y^K)||^2] \leq \epsilon_Y$ and $E[||{W_U}(R^K - U^K)||^2] \leq \epsilon_U$, for some desired maximum distortion levels ${\epsilon _Y} \in {\mathbb{R}_{>0}}$ and ${\epsilon _U} \in {\mathbb{R}_{>0}}$ and weights ${W_Y} \in {\mathbb{R}^{{n_y} \times {n_y}}}$ and ${W_U} \in {\mathbb{R}^{{n_u} \times {n_u}}}$, by designing $G(k)$, ${\Sigma^V(k)}$, and ${\Sigma^H(k)}$ of the distorting mechanisms \eqref{eq2}. In what follows, we formally present the optimization problem we seek to address.

\textbf{Problem 1} Given the system dynamics \eqref{eq1}, time horizon $K\in \mathbb{N}$, desired maximum distortion levels ${\epsilon _Y} \in {\mathbb{R}_{>0}}$ and ${\epsilon _U} \in {\mathbb{R}_{>0}}$, weighting matrices ${W_Y} \in {\mathbb{R}^{{n_y} \times {n_y}}}$ and ${W_U} \in {\mathbb{R}^{{n_u} \times {n_u}}}$, private output sequence $S\left( k \right)$, $k \in \mathcal{K}={1,...,K}$, and the distorting mechanism \eqref{eq2}, find the distorting sequences, $G(k)$, ${\Sigma^V(k)}$, and ${\Sigma^H(k)}$, solution of the following optimization problem:
\begin{equation} \label{problem1}
\left\{\begin{aligned}
	&\min_{G(k), {\Sigma^V(k)}, {\Sigma^H(k)}, k \in \mathcal{K}}\ I[S^K;Z^K] - h[{H}^{K}],\\[1mm]
    &\hspace{4mm}\text{s.t. } \left\{\begin{aligned}
    &E[||{W_Y}(Z^K - Y^K)||^2] \leq \epsilon_Y ,\\
    &E[||{W_U}(R^K - U^K)||^2] \leq \epsilon_U
    \end{aligned}\right. \\[1mm]
    &\hspace{12mm}\text{and } (V^K,H^K,Y^K) \text{ mutually independent}.
\end{aligned}\right.
\end{equation}

\section{Solution to Problem 1}
To solve Problem 1, we first need to write the cost function and constraints in terms of the design variables.

\subsection{Cost Function: Formulation and Convexity}
The differential entropy $h[{H}^{K}]$ is fully characterized by the covariance matrix of ${H}^{K}$, $\Sigma^H_K := E[H_K H_K^\top]$. Hence, by Definition 2, $h[{H}^{K}]$ is given by
\begin{equation}\label{hUprim}
h[{H}^{K}] = \frac{1}{2}\log \det \left( \Sigma^H_K \right) + \frac{Kn_u}{2} + \frac{Kn_u}{2}\log(2\pi).
\end{equation}

The mutual information $I\left[ {{S^K};{Z^K}} \right]$ can be written in terms of differential entropies as $I[S^K;Z^K] = h[S^K] + h[Z^K] - h[S^K,Z^K]$, see Definition \ref{mutual_info}. Moreover, these entropies are fully characterized by the covariance matrices of the corresponding random vectors, see Definition 2. So, to characterize $I\left[ {{S^K};{Z^K}} \right]$ in terms of $G(k)$ and $\Sigma^V(k)$, $k \in \mathcal{K}$, we need to write the covariance matrices of $S^K$ and $Z^K$, and the joint covariance of $(S^K,Z^K)$ in terms of them. By lifting the system dynamics \eqref{eq1} over $\{1,\ldots,K\}$, we can write the stacked vector $((Z^K)^\top,(S^K)^\top)^\top \in \mathbb{R}^{K(n_y + n_s)}$ as

\begin{align}\label{stackedXY}
&\begin{bmatrix} Z^K \\ S^K \end{bmatrix} = \begin{bmatrix} {{\tilde{G}}_K} \tilde{C}_K  \\ \tilde{D}_K \end{bmatrix}F_K X(1) + \begin{bmatrix} {{\tilde{G}}_K} \tilde{C}_K  \\ \tilde{D}_K \end{bmatrix} J_K T^{K-1}\\
&\hspace{15mm} + \begin{bmatrix} {{\tilde{G}}_K} \tilde{C}_K  \\ \tilde{D}_K \end{bmatrix} L_K U^{K-1} + \begin{bmatrix} I \\ \mathbf{0} \end{bmatrix}({{\tilde{G}}_K} W^{K}+V^{K}), \notag
\end{align}
with stacked matrices $L_K := J_K(I_{K-1} \otimes B)$, $\tilde{C}_K := I_K \otimes C$, $\tilde{D}_K := I_K \otimes D$, ${{\tilde{G}}_K} := \text{diag}\left[ {{G_1},{G_2}, \ldots ,{G_K}} \right]$, and
\begin{equation}\label{stackedZ}
\left\{ \begin{aligned}
  F_K &:= \begin{bmatrix} I & A^\top & \hdots & (A^\top)^{K-1}  \end{bmatrix}^\top,\\
  J_K &:= \begin{bmatrix} \mathbf{0} & \mathbf{0} & \mathbf{0} & \cdots & \mathbf{0} \\ I & \mathbf{0} & \mathbf{0} & \cdots & \mathbf{0} \\ A & I & \mathbf{0} & \cdots & \mathbf{0} \\ \vdots & \vdots & \vdots & \ddots & \vdots \\[1mm] (A)^{K-2} & (A)^{K-3} & (A)^{K-4} & \cdots & I  \end{bmatrix}.
\end{aligned} \right.
\end{equation}
Let ${{\Sigma }^V_K} \in \mathbb{R}^{K n_y \times K n_y}$ denote the \emph{non-diagonal} covariance matrix of the stacked additive dependent vector $V^K$. Note that matrices $\tilde{G}_K$ and ${{\Sigma}^V_K}$ contain all the distorting variables of the output mechanism, $(G(k),\Sigma^V(k))$, $k \in \mathcal{K}$. In the next lemma, we give a closed-form expression of the joint density of $(S^K,Z^K)$ (which we will need to write $I[Z^K;S^K]$ in terms of the design variables).

\begin{lemma}\label{stackedDist}
\[ \begin{psmallmatrix} Z^K \\ S^K  \end{psmallmatrix} \sim \mathcal{N}\left[\mu^{Z,S}_K,\Sigma^{Z,S}_K\right],\] with mean $\mu^{Z,S}_K \in \mathbb{R}^{K(n_s + n_y)}$ and covariance matrix $\Sigma^{Z,S}_K \in \mathbb{R}^{K(n_s + n_y) \times K(n_s + n_y)}$, $\Sigma^{Z,S}_K>0$\emph{:}
\begin{eqnarray}\label{muZS}
\mu^{Z,S}_K = \begin{bsmallmatrix} {{\tilde{G}}_K} \tilde{C}_K  \\ \tilde{D}_K \end{bsmallmatrix}F_K \mu^X_1 + \begin{bsmallmatrix} {{\tilde{G}}_K} \tilde{C}_K  \\ \tilde{D}_K \end{bsmallmatrix} L_K U^{K-1},
\end{eqnarray}
\begin{equation}\label{stackeconmatrix3}
\begingroup
\renewcommand*{\arraycolsep}{2pt}
\begin{aligned}
&\Sigma^{Z,S}_K := \begin{bsmallmatrix} {{\tilde{G}}_K} \\ \mathbf{0} \end{bsmallmatrix}(I_K \otimes \Sigma^W) \begin{bsmallmatrix} {{\tilde{G}}_K} \\ \mathbf{0} \end{bsmallmatrix}^\top + \begin{bsmallmatrix} I \\ \mathbf{0} \end{bsmallmatrix}{{\Sigma }^V_K}\begin{bsmallmatrix} I \\ \mathbf{0} \end{bsmallmatrix}^\top\\[1mm]
&+ \begin{bsmallmatrix} {{\tilde{G}}_K} \tilde{C}_K  \\ \tilde{D}_K \end{bsmallmatrix} F_K \Sigma^X_1 F_K^\top \begin{bsmallmatrix} {{\tilde{G}}_K} \tilde{C}_K  \\ \tilde{D}_K \end{bsmallmatrix}^\top\\[1mm]
&+ \begin{bsmallmatrix} {{\tilde{G}}_K} \tilde{C}_K  \\ \tilde{D}_K \end{bsmallmatrix}J_K (I_{K-1} \otimes \Sigma^T)J_K^\top \begin{bsmallmatrix} {{\tilde{G}}_K} \tilde{C}_K  \\ \tilde{D}_K \end{bsmallmatrix}^\top.
\end{aligned}
\endgroup
\end{equation}
\end{lemma}
\emph{\textbf{Proof}}: To simplify notation, we introduce the stacked vector ${\Theta ^K} := {( {{{( {{Z^K}})}^\top},{{( {{S^K}})}^\top}} )^\top}$. By assumption, the initial condition $X(1)$, and the processes, $T(k)$, $W(k)$, and $V(k)$, $k \in \mathbb{N}$, are mutually independent, and $X(1) \sim \mathcal{N} [ {\mu _1^X,\Sigma _1^X} ]$, $T\left( k \right) \sim \mathcal{N} [ {\mathbf{0},\Sigma^T} ]$, $W(k) \sim \mathcal{N} [{\mathbf{0},\Sigma^W} ]$, $V(k) \sim \mathcal{N} [ {\mathbf{0},\Sigma^V} ]$, for some positive definite covariance matrices $\Sigma _1^X$, $\Sigma^T$, $\Sigma^W$, and $\Sigma^V$. Then, see \cite{Ross} for details, we have ${L_1}X\left( 1 \right) \sim \mathcal{N} [ {{L_1}\mu _1^X,{L_1}\Sigma _1^XL_1^\top} ]$, ${L_2}T^{K-1} \sim \mathcal{N} [\mathbf{0}, {{L_2}({I_{K - 1}} \otimes {\sum ^T}){L_2}^T} ]$, ${L_3}W^K \sim \mathcal{N} [\mathbf{0}, {{L_3}({I_K} \otimes {\sum ^W}){L_3}^\top} ]$, ${L_4}V^K \sim \mathcal{N} [\mathbf{0}, {{L_4}{{\Sigma }^V_K}{L_4}^\top} ]$, for any deterministic matrices $L_j$,
$j = 1, 2, 3, 4$, of appropriate dimensions. It follows that the stacked vector ${\Theta ^K}$ given in \eqref{stackedXY} is the sum of a deterministic vector, ${( {({{\tilde{G}}_K} \tilde{C}_K)^\top,\tilde{D}_K^T} )^\top}{L_K}{U^{K - 1}}$, and four independent normally distributed vectors. Therefore, ${\Theta ^K}$ follows a multivariate normal distribution with $E[ {{\Theta ^K}} ] = \mu^{Z,S}_K$ as in \eqref{muZS}.
By inspection, using the expression of ${\Theta ^K}$ in \eqref{stackedXY}, mutual independence among $X(1)$, $T(k)$, $W(k)$, and $V(k)$, $k \in N$, and the definition of $\Sigma _1^X$, $\Sigma _1^X = E[ {\left( {X(1) - \mu _1^X} \right){{\left( {X(1) - \mu _1^X} \right)}^\top}} ]$, it can be verified that the covariance matrix of ${\Theta ^K}$, $E[ {( {{\Theta ^K} - E( {{\Theta ^K}} )} ){{( {{\Theta ^K} - E( {{\Theta ^K}} )} )}^\top}} ]$, is given by $\Sigma^{Z,S}_K$ in \eqref{stackeconmatrix3}. It remains to prove that the distribution of ${\Theta ^K}$ is not degenerate, that is, $\Sigma^{Z,S}_K > 0$. Note that $\Sigma^{Z,S}_K$ in \eqref{stackeconmatrix3} can be written as
\begin{align}
\Sigma _K^{Z,S} = \left[ {\begin{array}{*{20}{c}}
\Sigma _K^{Z} & { \tilde{G}_K  \tilde{C}_K Q  \tilde{D}_K^\top}\\
{ \tilde{D}_K Q  \tilde{C}_K^\top  {{\tilde{G}}_K}^\top}&{ \tilde{D}_K Q  \tilde{D}_K^\top}
\end{array}} \right],\label{CovZS}
\end{align}
\begin{eqnarray}
\Sigma _K^{Z} = {\tilde{G}_K \tilde{C}_K Q \tilde{C}_K^\top  {{\tilde{G}}_K}^\top +  {{\tilde{G}}_K} (I_K \otimes \Sigma^W) {{\tilde{G}}_K}^\top + {{\Sigma }^V_K}},
\end{eqnarray}
with $Q = {F_K}\Sigma _1^XF_K^\top + {J_K}( {{I_{K - 1}} \otimes {\sum ^T}} )J_K^\top$. A necessary condition for the block matrix $\Sigma _K^{Z,S}$ in \eqref{CovZS} to be positive definite is that the diagonal blocks are positive definite \cite{Horn}.
The left-upper block is positive definite because by assumption ${{\Sigma }^V_K} > 0$. The right-lower block is positive definite if $\tilde{D}_K$ is full row rank and $Q$ is positive definite. Because $D$ is full row rank by assumption, matrix $\tilde{D}_K = \left( {{I_K} \otimes D} \right)$ is also full row rank [\cite{Horn2}, Theorem 4.2.15].
Note that $Q$ can be factored as follows
\begin{eqnarray}\label{Q}
Q = \left[ {\begin{array}{*{20}{c}}
{{F_K}}&{{J_K}}
\end{array}} \right]\underbrace {\left[ {\begin{array}{*{20}{c}}
{\Sigma _1^X}&0\\
0&{{I_{K - 1}} \otimes {\Sigma ^T}}
\end{array}} \right]}_{Q'}\left[ {\begin{array}{*{20}{c}}
{{F_K}}\\
{{J_K}}
\end{array}} \right],
\end{eqnarray}
That is, $Q$ is a linear transformation of the block diagonal matrix ${Q'}$ above. By inspection, it can be verified that matrix $P = [F_K \hspace{1mm} J_K]$, see \eqref{stackedZ}, is lower triangular with identity matrices on the diagonal; thus, $P$ is invertible. It follows that $Q = PQ'P^\top$ is a congruence transformation of ${Q'}$ \cite{boyd1994linear}. Hence, $Q$ is positive definite if and only if the block diagonal matrices of ${Q'}$ are positive definite \cite{boyd1994linear}. Matrices ${\Sigma _1^X}$ and ${\Sigma^T}$ are positive definite by assumption (which implies ${{I_{K - 1}} \otimes {\Sigma ^T}}>0$), and thus we can conclude that $Q>0$, which implies ${\tilde{D}_K Q  \tilde{D}_K^\top}>0$ because $\tilde{D}_K$ is full row rank. Necessary and sufficient conditions for $\Sigma _K^{Z,S}>0$ are ${\tilde{D}_K Q  \tilde{D}_K^\top}>0$ which we have already proved) and that the Schur complement of block ${\tilde{D}_K Q  \tilde{D}_K^\top}$ of $\Sigma _K^{Z,S}$, denoted as $\Sigma _K^{Z,S} / ({\tilde{D}_K Q  \tilde{D}_K^\top})$, is positive definite [\cite{zhang2006schur}, Theorem 1.12]. This Schur complement is given by
\begin{align}
&\Sigma_K^{Z,S} / ({\tilde{D}_K Q  \tilde{D}_K^\top}) =
{{\tilde{G}}_K} (I_K \otimes \Sigma^W) {{\tilde{G}}_K}^\top + {{\Sigma }^V_K} +\\ \nonumber
&{{\tilde{G}}_K} \tilde{C}_K ( {Q - Q \tilde{D}_K^\top ({\tilde{D}_K Q \tilde{D}_K^\top} )^{-1}\tilde{D}_K Q}) \tilde{C}_K^\top {{\tilde{G}}_K}^\top.
\end{align}
Since matrix ${{\tilde{G}}_K} (I_K \otimes \Sigma^W) {{\tilde{G}}_K}^\top + {{\Sigma }^V_K}$ is positive definite, a sufficient condition for $\Sigma _K^{Z,S} / ({\tilde{D}_K Q  \tilde{D}_K^\top})>0$ is
\begin{eqnarray}
Q'' := ( {Q - Q \tilde{D}_K^\top ( {\tilde{D}_K Q \tilde{D}_K^\top} )^{-1}\tilde{D}_K Q} )>0.
\end{eqnarray}
Regarding ${Q''}$ as the Schur complement of a higher dimensional matrix ${Q'''}$, we can conclude that:
\begin{eqnarray}
\begin{array}{l}
Q'' \ge 0 \Leftrightarrow Q''' = \left[ {\begin{array}{*{20}{c}}
Q&{Q\tilde{D}_K^\top}\\
{\tilde{D}_K Q}&{\tilde{D}_K Q \tilde{D}_K^\top}
\end{array}} \right]\\
 = \left[ {\begin{array}{*{20}{c}}
Q\\
{\tilde{D}_K Q}
\end{array}} \right]{Q^{ - 1}}\left[ {\begin{array}{*{20}{c}}
Q&{Q\tilde{D}_K^\top}
\end{array}} \right] \ge 0,
\end{array}
\end{eqnarray}
which is trivially true because $Q^{-1}$ is positive definite since $Q>0$. Hence, ${\tilde{D}_K Q \tilde{D}_K^\top}$ and $\Sigma _K^{Z,S} / ({\tilde{D}_K Q  \tilde{D}_K^\top})$ are both positive definite, and thus $\Sigma _K^{Z,S}>0$.
\hfill $\blacksquare$

To obtain the densities of $Z^K$ and $S^K$, we just marginalize (see \cite{Ross}) their joint density in Lemma 1 over $S_K$ and $Z_K$.

\begin{corollary} \label{corollary1}
$Z^{K} \sim \mathcal{N} [\mu^Z_K, \Sigma^Z_K]$ and
$S^{K} \sim \mathcal{N}[\mu^S_K,\Sigma^S_K]$\emph{:}
\begin{align*}
&\mu^Z_K = {{\tilde{G}}_K} \tilde{C}_K F_K \mu^X_1 + {{\tilde{G}}_K} \tilde{C}_K L_K U^{K-1},\\
&\mu^S_K = \tilde{D}_KF_K \mu^X_1 + \tilde{D}_KL_KU^{K-1},\\
&\Sigma^Z_K = \begin{pmatrix} I_{n_y}  \\ \mathbf{0} \end{pmatrix}^\top \Sigma^{Z,S}_K \begin{pmatrix} I_{n_y} \\ 0  \end{pmatrix} \in \mathbb{R}^{Kn_y \times Kn_y},\\
&\Sigma^S_K = \begin{pmatrix} \mathbf{0}  \\ I_{n_s} \end{pmatrix}^\top \Sigma^{Z,S}_K \begin{pmatrix} \mathbf{0} \\ I_{n_s}  \end{pmatrix} \in \mathbb{R}^{Kn_s \times Kn_s}.
\end{align*}
\end{corollary}

At this point, we have the three covariance matrices, $(\Sigma^Z_K,\Sigma^S_K,\Sigma^{SZ}_K)$, that we need to compute the mutual information $I\left[ {{S^K};{Z^K}} \right]$. Note that these matrices are functions of our design variables $\tilde{G}_K$ and $\Sigma^V_K$:
\begin{align}
&\Sigma _K^{Z,S} = \begin{pmatrix} \Sigma _K^{Z} & (\Sigma _K^{SZ})^\top \\ \Sigma _K^{SZ} & \Sigma _K^{S} \end{pmatrix},\label{SigmaS_Z}\\[1mm]
&\Sigma_K^{Z} = {{{\tilde{G}}_K} \tilde{C}_K Q \tilde{C}_K^\top  {{\tilde{G}}_K}^\top +  {{\tilde{G}}_K} (I_K \otimes \Sigma^W) {{\tilde{G}}_K}^\top + {{\Sigma }^V_K}},\label{eq3g}\\[1mm]
&\Sigma^S_K = { \tilde{D}_K Q  \tilde{D}_K^\top},\label{eq3h}\\[1mm]
&\Sigma _K^{SZ} = { {{\tilde{G}}_K}  \tilde{C}_K Q  \tilde{D}_K^\top},\label{SigmaSZ}
\end{align}
with $Q$ as in \eqref{Q} independent of $\tilde{G}_K$ and $\Sigma^V_K$. Before we write the cost in terms of these matrices, we notice that $\Sigma^V_K$ only appears in the expression for $\Sigma^Z_K$ in \eqref{eq3g}. Moreover, given $(\tilde{G}_K,\Sigma^Z_K)$, matrix $\Sigma^V_K$ is fully determined and viceversa. That is, $(\tilde{G}_K,\Sigma^V_K) \rightarrow (\tilde{G}_K,\Sigma^Z_K)$ is an invertible transformation. Therefore, we can pose both cost and constraints in terms of either $\Sigma^V_K$ or $\Sigma^Z_K$. Casting the problem in terms of $\Sigma^Z_K$ allows us to write linear distortion constraints and a convex cost function. Hereafter, we pose the problem in terms of $(\tilde{G}_K,\Sigma^Z_K)$. Once we have found optimal $(\tilde{G}_K,\Sigma^Z_K)$, we simply extract the optimal $\Sigma^V_K$ using \eqref{eq3g}. The first constraint that we need to enforce is that the extracted  $\Sigma^V_K$ is always positive definite (as it is a covariance matrix). Having $\Sigma^Z_K > \mathbf{0}$ does not necessary imply $\Sigma^V_K > \mathbf{0}$. From \eqref{eq3g}, it is easy to verify that $\Sigma^V_K>\mathbf{0}$ if and only if $\Sigma_K^{Z} - {{\tilde{G}}_K} (\tilde{C}_K Q \tilde{C}_K^\top + (I_K \otimes \Sigma^W)) {{\tilde{G}}_K}^\top  > \mathbf{0}$. Using standard Schur complement properties \cite{zhang2006schur}, the latter nonlinear inequality can be rewritten as a higher-dimensional linear matrix inequality in $\Sigma^Z_K$ and $\tilde{G}_K$ as follows:
\begin{eqnarray}\label{SigmaV_pos_def}
\left[ {\begin{array}{*{20}{c}}
\Sigma _K^Z    &    {{\tilde{G}}_K} \\
{{\tilde{G}}_K}^\top      &    (\tilde{C}_K Q \tilde{C}_K^\top + (I_K \otimes \Sigma^W))^{-1}
\end{array}} \right] > 0.
\end{eqnarray}
We use inequality \eqref{SigmaV_pos_def} later when we solve the complete optimization problem to enforce that the optimal $\Sigma^Z_K$ and $\tilde{G}_K$ lead to a positive definite $\Sigma^V_K$.

Finally, given $(\Sigma^Z_K,\Sigma^S_K,\Sigma^{SZ}_K)$ in \eqref{SigmaS_Z}-\eqref{SigmaSZ}, by Definition 2, we can write $I[Z^K;S^K]$ as follows:
\begin{subequations}
\begin{align}
&I[Z^K;S^K] = h\left( S^K \right) + h\left( Z^K \right) - h(Z^K,S^K)\label{eq3a}\\[1mm]
& = \frac{1}{2} \log \frac{{\det \left( {\Sigma _K^S} \right)\det \left( {\Sigma _K^Z} \right)}}{{\det \left( {\Sigma _K^{Z,S}} \right)}}\label{eq3d}\\
& = \frac{1}{2} \log{\det \left( {\Sigma _K^S} \right)} - \frac{1}{2} \log{\det  {\left( {\Sigma} _K^S - {{\Sigma} _K^{SZ}}^\top {{\Sigma} _K^ Z}^{-1} {\Sigma} _K^{SZ} \right),}}\label{eq3f}
\end{align}
\end{subequations}
where \eqref{eq3d}-\eqref{eq3f} follow from standard determinant and logarithm formulas. In the following lemma, we prove that minimizing the cost function $I[S^K;Z^K] - h[{H}^K]$ using $({{\tilde{G}}_K},{{\Sigma }^Z_K},{{\Sigma }^H_K})$ as optimization variables is equivalent to solving a convex program subject to some Linear Matrix Inequalities (LMI) constraints.

\begin{lemma}\label{mutualinformationcov}
Minimizing $I[S^K;Z^K] - h[{H}^K]$ is equivalent to solving the following convex program:
\begin{eqnarray}
\left\{\begin{aligned}
	&\min_{\Sigma^H_K, {\Pi _K},\Sigma_K^{Z},{{\tilde{G}}_K}}\
      -\log \det \left( \Sigma^H_K \right) - \log{\det \left({\Pi _K} \right)} \label{finalcost_program}\\[1mm]
    &\hspace{4mm}\text{\emph{s.t. }} \Pi _K \geq \mathbf{0}, \begin{bmatrix}
\Sigma _K^S - \Pi _K & (\Sigma_K^{SZ})^\top\\
\Sigma_K^{SZ} & \Sigma_K^Z
\end{bmatrix} \geq 0.
\end{aligned}\right.
\end{eqnarray}
\end{lemma}
\textbf{\emph{Proof}}: {(a)} For any positive definite matrix $\Sigma$, the function $-\log\det(\Sigma)$ is convex in $\Sigma$ \cite{boyd2004convex}. It follows that $-h[{H}^K]$ in \eqref{hUprim} is a convex function of $\Sigma^H_K$. Minimizing $-h[{H}^K]$ amounts to minimizing $-\log \det \left( \Sigma^H_K \right)$ as all other terms in \eqref{hUprim} are constants. {(b)} Next, consider the expression for $I[Z^K;S^K]$ in \eqref{eq3f}. Due to monotonicity of the determinant function and the fact that $\Sigma _K^S$ is independent of the design variables, minimizing \eqref{eq3f} is equivalent to
\begin{eqnarray}
\left\{\begin{aligned}
	&\min_{{\Pi _K},\Sigma_K^{Z},{{\tilde{G}}_K}}\
      - \log{\det \left({\Pi _K} \right)} \label{epigraphcost}\\[1mm]
    &\hspace{4mm}\text{s.t. } \mathbf{0} < {\Pi _K} \le  {\left( {\Sigma} _K^S - {{\Sigma} _K^{SZ}}^\top {{\Sigma} _K^ Z}^{-1} {\Sigma} _K^{SZ} \right).} \label{inequalityofcost}
\end{aligned}\right.
\end{eqnarray}
{(c)} The inequality term in \eqref{inequalityofcost} can be rewritten using Schur complements properties  \cite{zhang2006schur} as
\begin{eqnarray}
&\left[ {\begin{array}{*{20}{c}}
{\Sigma _K^S} - {\Pi _K} & {{\Sigma} _K^{SZ}}^\top\\
{{\Sigma} _K^{SZ}}&{{\Sigma} _K^ Z}
\end{array}} \right] \ge 0 \, {,{\Pi _K} > 0}. \label{finalcost}
\end{eqnarray}
Because $\Sigma_K^{SZ}$ is a linear function of $\tilde{G}_K$ (see \eqref{SigmaSZ}), \eqref{finalcost} are two LMI constraints in $\Pi _K$ and $\tilde{G}_K$. Combining {(a)-(c)}, we can conclude that minimizing $I[S^K;Z^K] - h[{H}^K]$ is equivalent to solving the convex program in \eqref{finalcost_program}. \hfill $\blacksquare$

By Lemma \ref{mutualinformationcov}, minimizing the cost in \eqref{problem1} is equivalent to solving the convex program in \eqref{finalcost_program}. Then, if the distortion metrics, $E[||{W_Y}(Z^K - Y^K)||^2]$ and $E[||{W_U}({R}^K - U^K)||^2]$, are convex in the decision variables (which is the topic of the next section), we can find optimal distorting mechanisms efficiently using off-the-shelf optimization algorithms.

\subsection{Distortion Constraints: Formulation and Convexity}

We start with the distortion in $Y^K$. By lifting the system dynamics \eqref{eq1} over $\{1,\ldots,K\}$, we can write the mean and covariance matrix of the stacked output $Y^K \in \mathbb{R}^{Kn_y}$ as:
\begin{align}
\mu^{Y}_K &= \tilde{C}_K F_K \mu^X_1 + \tilde{C}_K L_K U^{K-1},  \label{eq4d}\\[1mm]
\Sigma^{Y}_K &= (I_{K} \otimes \Sigma^W) + \tilde{C}_K Q \tilde{C}_K^\top, \label{eq4e}
\end{align}
with matrices $L_K = J_K(I_{K-1} \otimes B)$, $\tilde{C}_K = I_K \otimes C$, $J_K$ in \eqref{stackedZ}, and $Q$ in \eqref{Q}.

\begin{lemma}\label{constraint}
$E[||{W_Y}(Z^K - Y^K)||^2]$ is a convex function of ${{\Sigma }^Z_K}$ and ${{\tilde{G}}_K}$, and can be written as follows:
\begin{align}\label{distortion}
&E[||{W_Y}(Z^K - Y^K)||^2] \hspace{20cm} \notag\\[1mm]
&\hspace{6mm} = \text{\emph{tr}}[{W_Y}^\top (\Sigma _K^Z + \Sigma _K^Y - 2\Sigma _K^Y\tilde{G}_K) {W_Y}]\notag\\[1mm]
& \hspace{14mm} + {{\mu _K^Y}^\top} ({{\tilde{G}}_K} - I)^\top {W_Y}^\top {W_Y} ({{\tilde{G}}_K} - I)  {\mu _K^Y}.
\end{align}
\end{lemma}
\textbf{\emph{Proof}}: The expectation of the quadratic form, $E[||{W_Y}(Z^K - Y^K)||^2]$, can be written (see \cite{seber2012linear} for details) in terms of the mean and covariance of the error $\Delta^K := Z^K - Y^K$:
\begin{align}
&E[||{W_Y}(Z^K - Y^K)||^2] = \text{tr}[\Sigma _K^\Delta] + (\mu _K^{\Delta})^\top \mu _K^\Delta \label{eq4b}.
\end{align}
with covariance $\Sigma _K^\Delta := E[(\Delta^K - \mu _K^{\Delta})(\Delta^K - \mu _K^{\Delta})^\top]$ and mean $\mu _K^{\Delta} := E[\Delta^K]$. Given the distortion mechanism \eqref{eq2}, we can write $Z^K = \tilde{G}_K Y^K + V^K$; hence, $\Delta^K = {W_Y}(({{\tilde{G}}_K} - I) Y^K + V^K)$, and because $Y^K$ and $V^K$ are independent, $\Sigma _K^\Delta = {W_Y}^\top ({{\tilde{G}}_K} - I)^\top \Sigma _K^Y ({{\tilde{G}}_K} - I) + {{\Sigma }^V_K}) {W_Y}$ and $\mu _K^\Delta = {W_Y} ({{\tilde{G}}_K} - I) \mu _K^Y$. Note that, because $Z^K = \tilde{G}_K Y^K + V^K$, $\Sigma^Z_K = {{\tilde{G}}_K}^\top \Sigma _K^Y {{\tilde{G}}_K} + {{\Sigma }^V_K}$; then, we can write $\Sigma _K^\Delta$ in terms of $\Sigma^Z_K$ as  $\Sigma _K^\Delta = {W_Y}^\top (\Sigma _K^Z + \Sigma _K^Y - {{\tilde{G}}_K}^\top \Sigma _K^Y - \Sigma _K^Y {{\tilde{G}}_K}) {W_Y}$. Up to this point, we have written both $\mu _K^{\Delta}$ and $\Sigma _K^\Delta$ in terms of the design variables, ${{\Sigma }^Z_K}$ and ${{\tilde{G}}_K}$. Using these expressions and \eqref{eq4b}, we can conclude \eqref{distortion}. \hfill $\blacksquare$

\begin{remark}
Note that the distortion metric \eqref{distortion} is linear in ${{\Sigma }^Z_K}$ and quadratic in ${{\tilde{G}}_K}$; consequently, the output distortion constraint in \eqref{problem1}, $E[||{W_Y}(Z^K - Y^K)||^2] \le \epsilon_Y$, is nonlinear in the design variables. However, we can find an equivalent linear constraint using standard Schur complement properties. It can be verified (see \emph{\cite{zhang2006schur}}, Theorem 1.12]) that $E[||{W_Y}(Z^K - Y^K)||^2] \le \epsilon_Y$ in \eqref{distortion} is equivalent to the following LMI in $\Sigma^Z_K$ and $\tilde{G}_K$:
\begin{equation}
\left\{
\begin{array}{ll}
\begin{bmatrix} \theta_Y & {{\mu _K^Y}^\top}({{\tilde{G}}_K} - I)^\top {W_Y}^\top \\ {W_Y} ({{\tilde{G}}_K} - I) {\mu _K^Y} & I \end{bmatrix} \ge 0,\\[6mm]
\theta_Y := \epsilon_Y - \text{\emph{tr}}[{W_Y}^\top (\Sigma _K^Z + \Sigma _K^Y - 2\Sigma _K^Y\tilde{G}_K) {W_Y}].
\end{array}
\right.
\end{equation}
\end{remark}

We move now to the distortion in $U^K$. The input distortion metric $E[||{W_U}(R^K - U^K)||^2] = E[||{W_U}H^K||^2]$ depends on the mean $\mu^H_K$ and covariance $\Sigma^H_K$ of $H^K$. Note that $E[||{W_U}H^K||^2]$ is again the expected value of a quadratic form. By construction ($H^K$ is a design vector), we have $\mu^H_K = \mathbf{0}$. Then, $E[||{W_U}H^K||^2]$ is simply given by $E[||{W_U}H^K||^2] = \text{tr}[{W_U}^\top {\Sigma }^H_K {W_U}]$ (see \cite{seber2012linear} for details), which is already linear in the design matrix $\Sigma^H_K$. We summarize this discussion in the following lemma.

\begin{lemma}\label{constraintU}
$E[||{W_U}(R^K - U^K)||^2] = E[||{W_U}H^K||^2]$ is a convex function of $\Sigma^H_K$ and can be written as follows:
\begin{eqnarray}
E[||{W_U}H^K||^2] = \text{\emph{tr}}[{W_U}^\top {\Sigma }^H_K {W_U}]. \label{distortionU}
\end{eqnarray}
\end{lemma}

By Lemma \ref{mutualinformationcov}, Lemma \ref{constraint}, and Lemma \ref{constraintU}, the cost function $I[S^K;Z^K] - h[{H}^K]$ and distortion constraints $E[||{W_Y}(Z^K - Y^K)||^2] \leq \epsilon_Y$ and $E[||{W_U}{H}^K||^2] \leq \epsilon_U$ can be written in terms of convex functions (programs) in the our optimization variables ${{\Sigma }^Z}$, ${{\tilde{G}}_K}$, and ${{\Sigma }^H_K}$.

In what follows, we pose the complete nonlinear convex program to solve Problem 1.

\begin{theorem}\label{th3}
Consider the system dynamics \eqref{eq1}, distorting mechanism \eqref{eq2}, time horizon $K \in \mathbb{N}$, desired input and output distortion levels ${\epsilon _U},{\epsilon _Y} \in {\mathbb{R}^+}$, input and output distortion weights, ${W_Y} \in {\mathbb{R}^{K {n_y} \times K {n_y}}}$ and ${W_U} \in {\mathbb{R}^{(K-1){n_u} \times (K-1){n_u}}}$, covariance matrices $\Sigma^{Z,S}_K$ and $\Sigma _K^{S}$ and cross-covariance $\Sigma _K^{SZ}$ \emph{(}given in \eqref{SigmaS_Z}-\eqref{SigmaSZ}\emph{)}, and the mean and covariance of $Y^K$, $\mu^{Y}_K$ and $\Sigma^{Y}_K$ \emph{(}given in \eqref{eq4d}-\eqref{eq4e}\emph{)}. Then, the optimization variables ${{\Sigma }^Z}$, ${{\tilde{G}}_K}$, and ${{\Sigma }^H_K}$ that minimize $I[S^K;Z^K] - h[{H}^K]$ subject to distortion constraints, $E[||{W_Y}(Z^K - Y^K)||^2] \leq \epsilon_Y$ and $E[||W_UH^K)||^2] \leq \epsilon_U$, can be found by solving the convex program in \eqref{eq:convex_optimization15}.
\end{theorem}
\emph{\textbf{Proof:}} The expressions for the cost and constraints and convexity (linearity) of them follow from Lemma \ref{mutualinformationcov}, Lemma \ref{constraint}, and Lemma \ref{constraintU}, Remark 1, and \eqref{SigmaV_pos_def}.  \hfill $\blacksquare$

\begin{table}
\noindent\rule{\hsize}{1pt}
\begin{equation} \label{eq:convex_optimization15}
\begin{aligned}
	&\min_{{\Pi _K}, \Sigma_K^{Z}, {{\tilde{G}}_K}, {{\Sigma }^H_K}}  -\log\det [ \Sigma^H_K ] - \log\det [\Pi_K ],\\[1mm]
    &\text{ s.t. }\left\{\begin{aligned}
    &{\Pi _K} > 0,\\ &\left[ {\begin{array}{*{20}{c}}
{\Sigma _K^S} - {\Pi _K} & ( {{\tilde{G}}_K}  \tilde{C}_K Q  \tilde{D}_K^\top)^\top\\
{{\tilde{G}}_K}  \tilde{C}_K Q  \tilde{D}_K^\top &{{\Sigma} _K^ Z}
\end{array}} \right] \ge 0, \\[1mm]
&\left[ {\begin{array}{*{20}{c}}
\theta_Y & ({W_Y} ({{\tilde{G}}_K} - I){\mu _K^Y})^\top\\
{W_Y} ({{\tilde{G}}_K} - I){\mu _K^Y}&I
\end{array}} \right] \ge 0, \\[1mm]
&\theta_Y = \epsilon_Y - \text{tr}[{W_Y}^\top (\Sigma _K^Z + \Sigma _K^Y - 2\Sigma _K^Y\tilde{G}_K) {W_Y}], \\[1mm]
&\epsilon_U - \text{tr}[{W_U}^\top {\Sigma }^H_K {W_U}] \ge 0,\\[1mm]
&\left[ {\begin{array}{*{20}{c}}
\Sigma _K^Z    &    {{\tilde{G}}_K} \\
{{\tilde{G}}_K}^\top      &    (\tilde{C}_K Q \tilde{C}_K^\top + (I_K \otimes \Sigma^W))^{-1}
\end{array}} \right] > 0 \\
&\Sigma _K^H > 0. \end{aligned}\right.
\end{aligned}
\end{equation}
\noindent\rule{\hsize}{1pt}
\end{table}

\section{Illustrative case study}
We illustrate the performance of our tools through a case study of a well-stirred chemical reactor with heat exchanger. This case study has been developed over the years as a benchmark example for control systems and fault detection, see, e.g., \cite{Wata,IET_CARLOS_JUSTIN} and references therein. The state, inputs, and output of the reactor are:
\begingroup\makeatletter\def\f@size{9.0}\check@mathfonts
\def\maketag@@@#1{\hbox{\m@th\normalsize\normalfont#1}}%
\begin{align*}
\left\{
\begin{array}{ll}
X(t) =  \begin{pmatrix} C_0\\T_0\\T_w\\T_m \end{pmatrix},
U(t) =  \begin{pmatrix} C_u\\T_u\\T_{w,u} \end{pmatrix},
Y(t) =  T_0,
\end{array}
\right.
\end{align*}\endgroup
where
\begingroup\makeatletter\def\f@size{9.0}\check@mathfonts
\def\maketag@@@#1{\hbox{\m@th\normalsize\normalfont#1}}%
\begin{align*}
\left\{
\begin{array}{ll}
C_0&: \text{Concentration of the chemical product},\\
T_0&: \text{Temperature of the product},\\
T_w&: \text{Temperature of the jacket water of heat exchanger},\\
T_m&: \text{Coolant temperature},\\
C_u&: \text{Inlet concentration of reactant},\\
T_u&: \text{Inlet temperature},\\
T_{w,u}&: \text{Coolant water inlet temperature}.
\end{array}
\right.
\end{align*}\endgroup
We use the discrete-time dynamics of the reactor introduced in \cite{murguia2021privacy} with the same noise and input signals for our simulation experiments.

As private output, we use the concentration of the chemical product; then, the matrix $D$ in \eqref{eq1} is given by the full row rank matrix $D = (1,0,0,0)$. The output of the system is the temperature of the product $T_0$, which could be monitored, e.g., for quality/safety reasons. Then, the aim of the privacy scheme is to hide the private state $C_0$,  which is the concentration of the reactant, as much as possible without distorting output signal temperature measurements and input signals excessively.
\begin{figure}[!htb]
  \centering
  \includegraphics[width=3.5in]{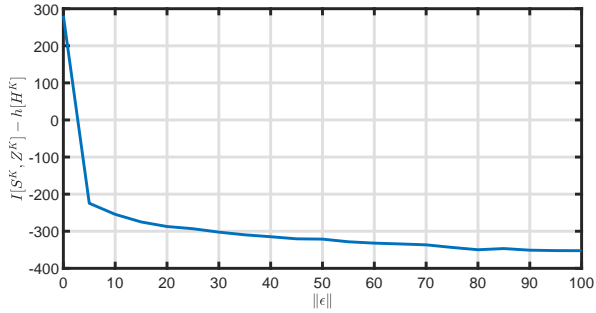}
  \caption{Evolution of the cost function for horizon $K=30$ and increasing $||\epsilon|| = ||(\epsilon_Y,\epsilon_U)||$.}\label{CostBasedEpsUEpsY}
\end{figure}
\begin{figure}[ht]
\centering
\begin{subfigure}{.5\textwidth}
  \centering
  \includegraphics[width=3.5in]{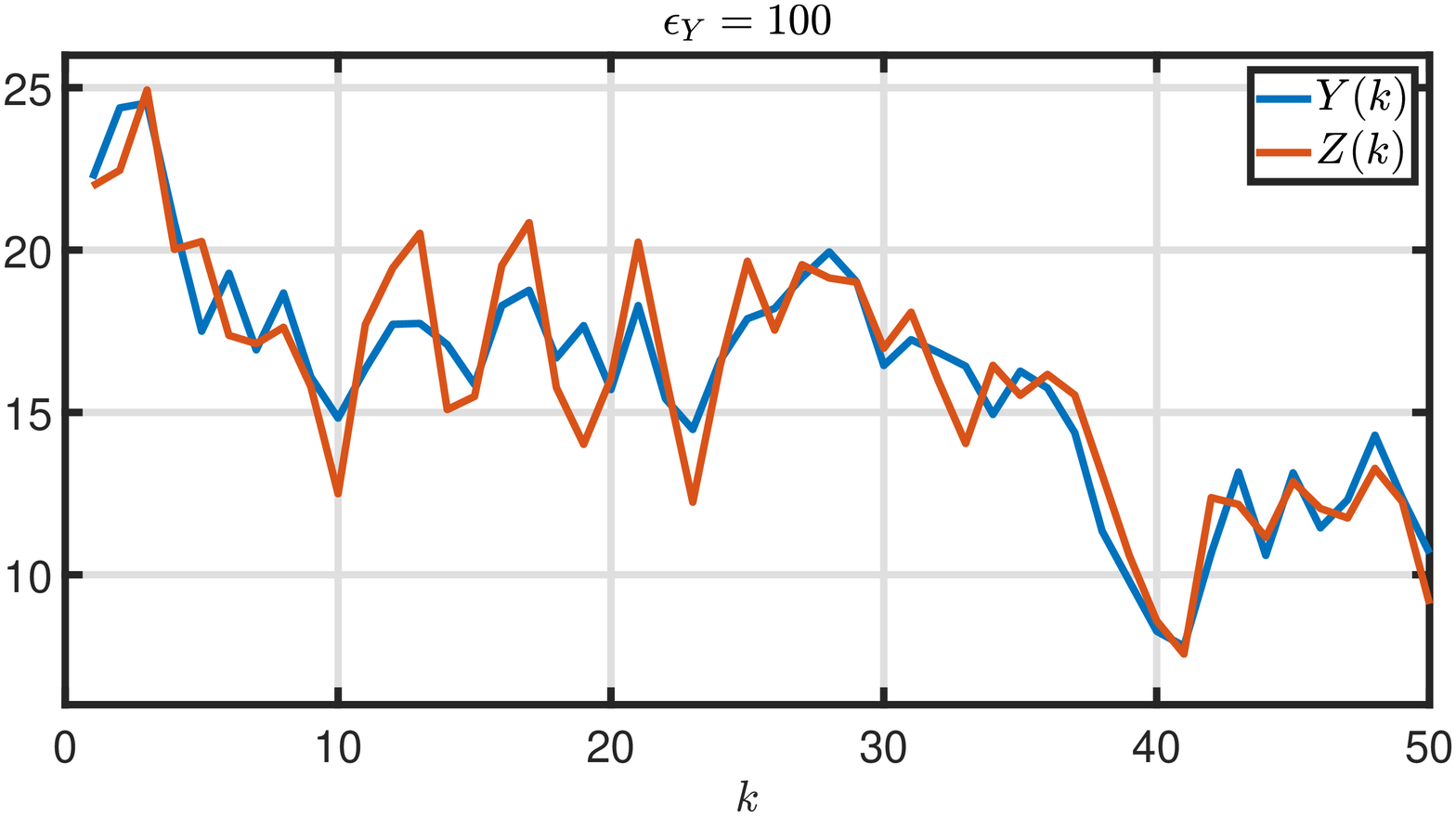}
  \label{fig:sub-first}
\end{subfigure}
\begin{subfigure}{.5\textwidth}
  \centering
  \includegraphics[width=3.5in]{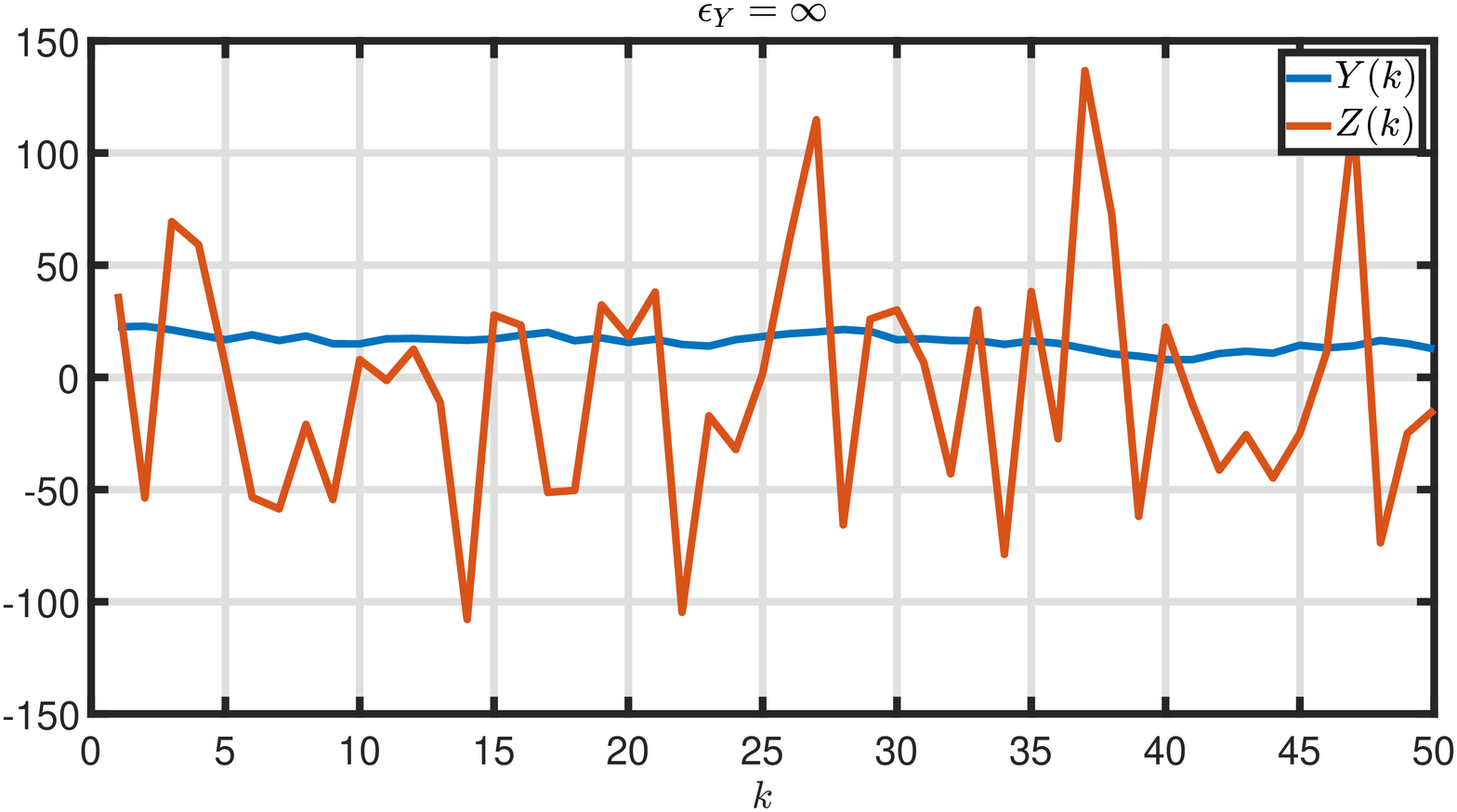}
  \label{fig:sub-second}
\end{subfigure}
\caption{Comparison between the measurement vector $Y(k)$ and the distorted measurements $Z(k)$ for two different distortion levels, $\epsilon_Y = 100, \infty$.}
\label{YZ}
\end{figure}

\begin{figure}[ht]
\centering
\begin{subfigure}{.5\textwidth}
  \centering
  \includegraphics[width=3.5in]{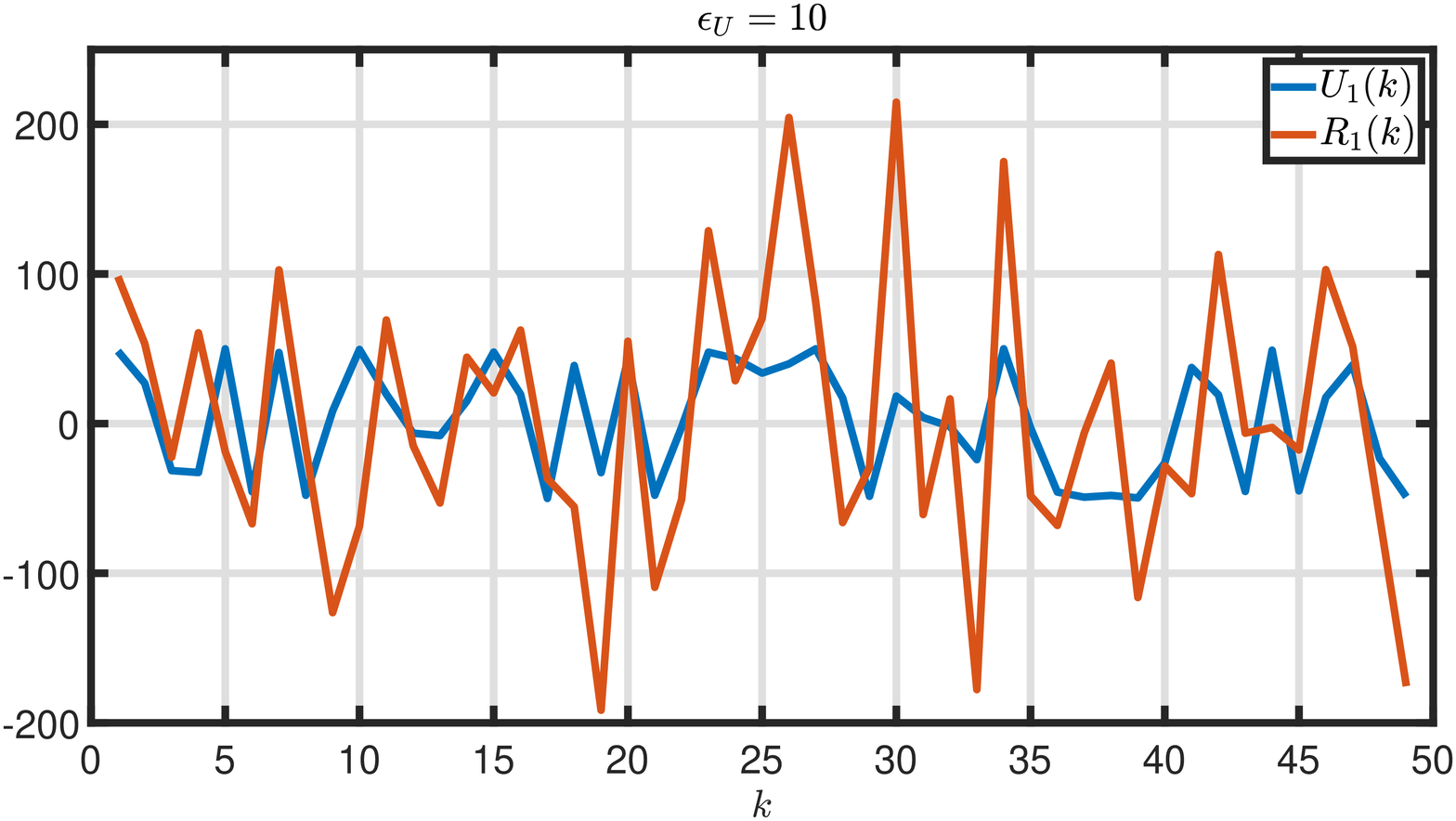}
  \label{fig:sub-first}
\end{subfigure}
\begin{subfigure}{.5\textwidth}
  \centering
  \includegraphics[width=3.5in]{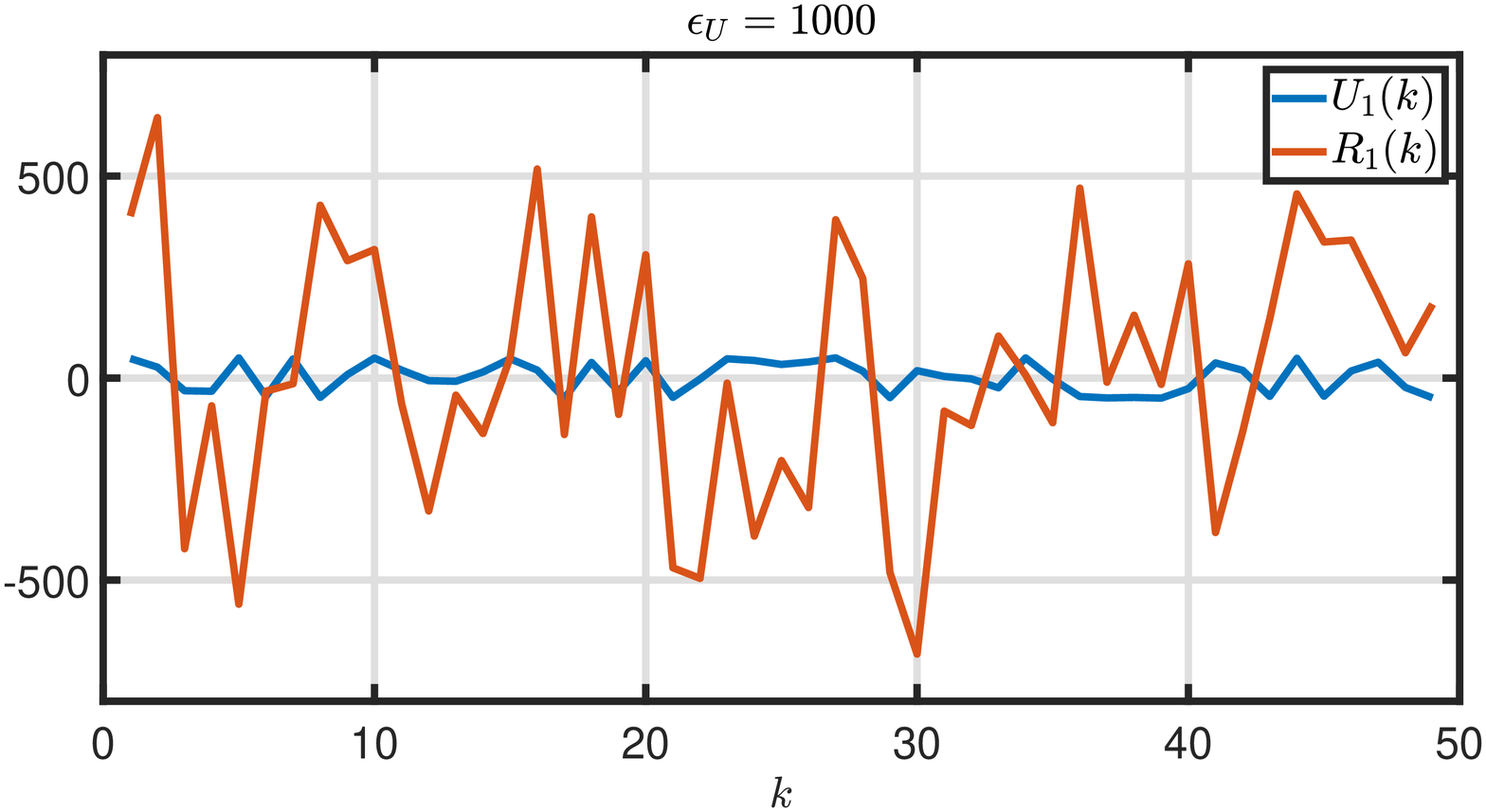}
  \label{fig:sub-second}
\end{subfigure}
\caption{Comparison between the first element of the input signal ${U}_1(k)$ and the first element of distorted input signal ${R}_1(k)$ for two different distortion levels, $\epsilon_U = 10, 1000$.}
\label{UUp}
\end{figure}
\begin{figure}[ht]
\centering
\begin{subfigure}{.5\textwidth}
  \centering
  \includegraphics[width=3.5in]{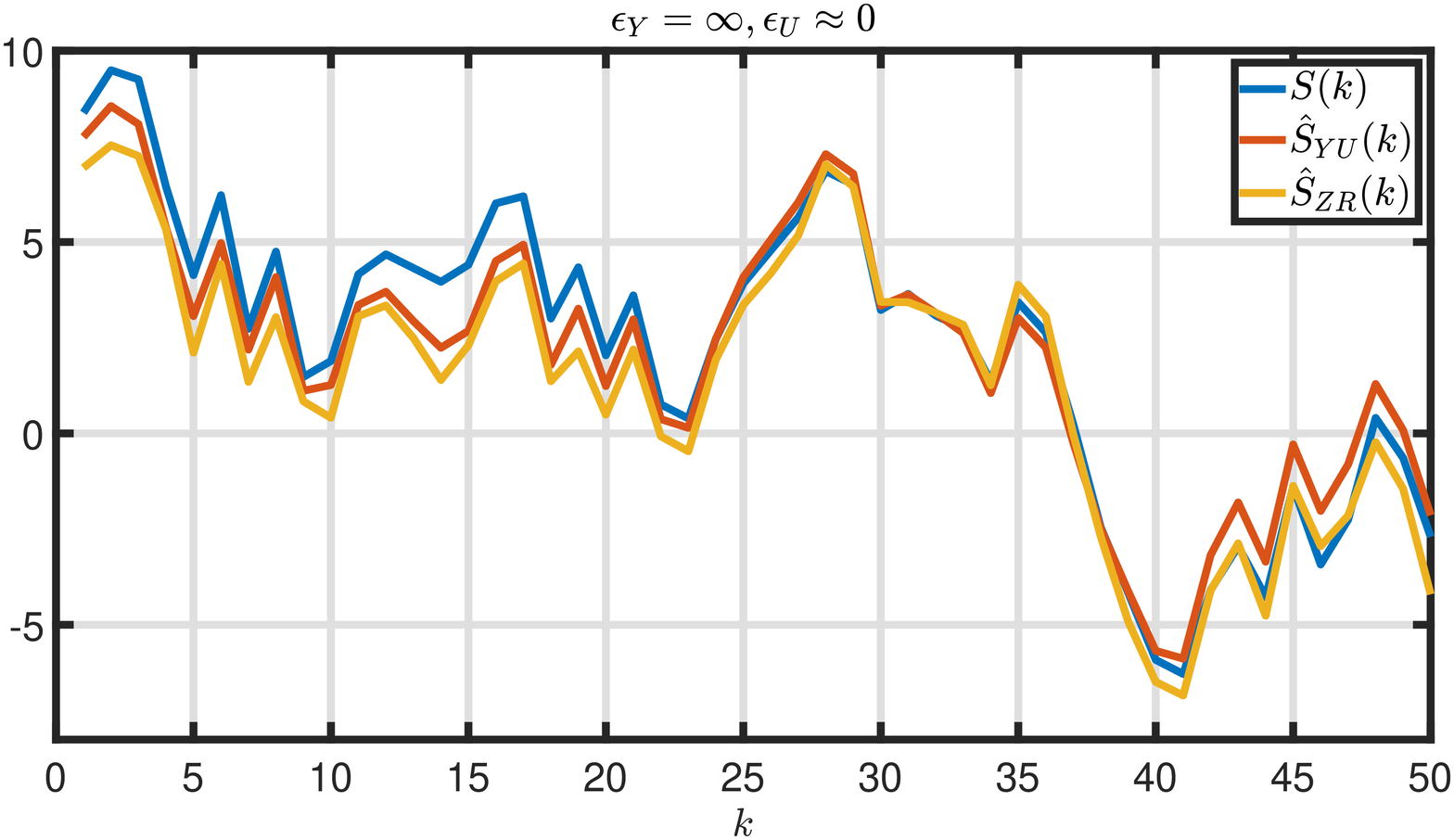}
  \label{S-sub-first}
\end{subfigure}
\begin{subfigure}{.5\textwidth}
  \centering
  \includegraphics[width=3.5in]{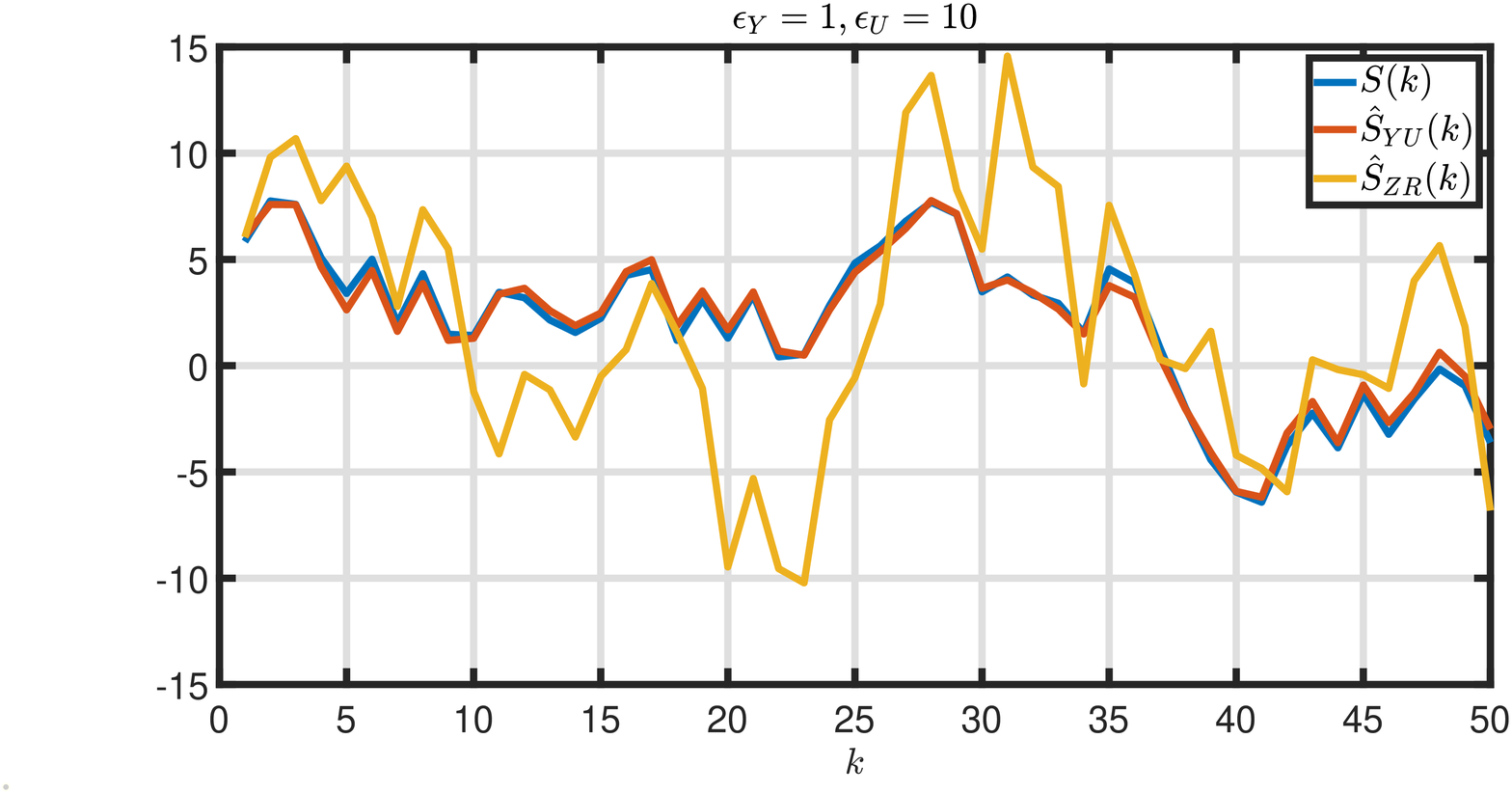}
  \label{S-sub-second}
\end{subfigure}
\begin{subfigure}{.5\textwidth}
  \centering
  \includegraphics[width=3.5in]{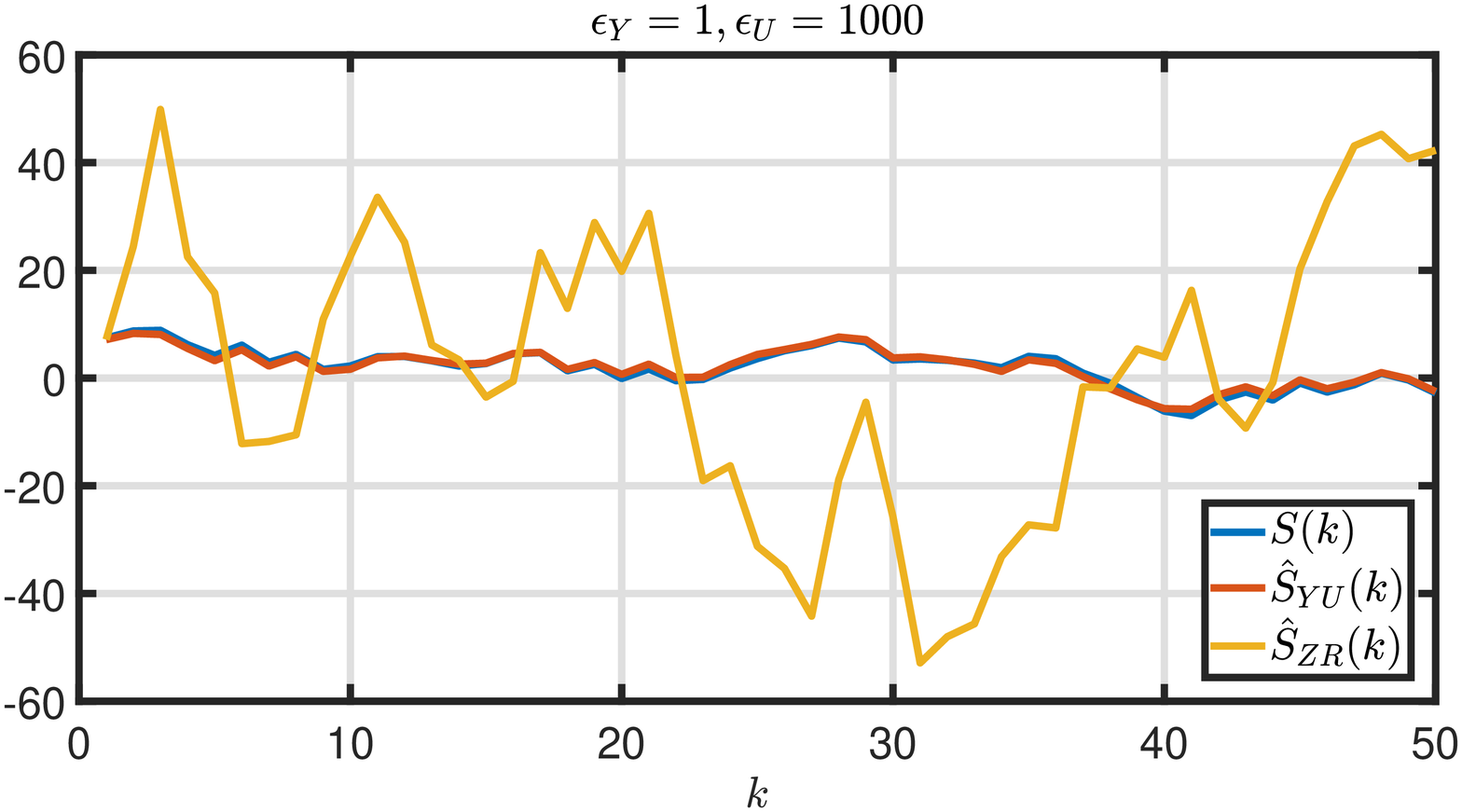}
  \label{S-sub-3rd}
\end{subfigure}
\caption{Comparison between the private output $S(k)$, the estimated private output without distorting $Y(k)$ and $U(k)$, ${\hat S_{YU}}(k)$, and the estimated private output given $Z(k)$ and $R(k)$, ${\hat S_{ZR}}(k)$, for horizon $K=50$, and three different levels of distortions, $\epsilon_Y$ and $\epsilon_U$.}
\label{estimatedS}
\end{figure}
We first show the effect of distortion levels $\epsilon_Y$ and $\epsilon_U$ in the cost function. Figure \ref{CostBasedEpsUEpsY} depicts the evolution of the optimal cost $I[S^K;Z^K] - h[H^K]$ for increasing $\epsilon = (\epsilon_Y,\epsilon_U)$ with time horizon $K=30$. As expected, the objective function decreases monotonically for increased maximum allowed distortion.
Furthermore, this figure illustrates that by increasing the distortion levels, the decreasing speed of the optimal cost function will be reduced. Therefore, we can design the distortion levels such that the information leakage is minimized without distorting measurement and input signals excessively.\\
The effect of the optimal distortion mechanisms is illustrated in Figure \ref{YZ} and Figure \ref{UUp} for different levels of distortion, where we contrast actual and distorted data.
Next, in Figure \ref{estimatedS}, we depict the stacked private output $S^K$, its MMSE estimate ${{\hat S}^K_{YU}}$ (see Definition 1) without distortion, i.e., the MMSE estimate of $S^K$ given $Y^K$ and $U^{K-1}$, and its MMSE estimate given the distorted vectors $Z^K$ and ${R}^{K-1}$, ${{\hat S}^K_{ZR}}$. We consider horizon $K=50$, and different levels of distortion, $(\epsilon_Y,\epsilon_U) = \{ (\infty,0), (1,10), (1,1000) \}$, to investigate the effect of each distortion level separately. $\epsilon_Y = \infty$ means that the optimization problem in \eqref{eq:convex_optimization15} is solved without considering the distortion constraint between $Z^K$ and $Y^K$. As can be seen in this figure, even after distorting the measurements without the distortion constraint, the stacked private vectors ${S^K}$ can be estimated accurately by a MMSE estimator. However, by randomizing $U^K$, we can prevent an accurate estimation. By increasing the distortion level $\epsilon_U$, the accuracy of the estimation decreases.\\

\section{Conclusions}
In this paper, for a class of stochastic dynamical systems, we have presented a detailed mathematical framework for synthesizing distorting mechanisms to minimize the information leakage induced by the use of public/unsecured communication networks. We have proposed a class of dependent Gaussian distorting mechanisms to randomize sensor measurements and input signals before transmission to prevent adversaries from accurately estimating the private part of the system state (a performance private output).\\
Furthermore, for the class of systems under study, we have fully characterized information-theoretic metrics (mutual information and differential entropy) to quantify the information between private outputs and disclosed data for a class of worst-case eavesdropping adversaries.\\
Finally, given the maximum level of distortion tolerated by a particular application, we have provided tools (in terms of convex programs) to design optimal (in terms of maximizing privacy) distorting mechanisms. We have presented simulation results to illustrate the performance of our tools.


\bibliographystyle{IEEEtran}
\bibliography{ifacconf32}
\end{document}